\documentstyle[twoside,espcrc2,fleqn]{article}

\def\refgen{\bibitem}

\def\h{\hbox to .5cm{\hfill}}
\def\hof{\hbox to .15cm{\hfill}}
\def\hi{\hbox to .2cm{\hfill}}
\def\htwo{\hbox to .2cm{\hfill}}
\def\htf{\hbox to .35cm{\hfill}}

\def\ha#1{\n\hbox to .6cm{\n {#1}\hfill}}
\def\hb#1{\indent\hbox to .7cm{\n {#1}\hfill}}
\def\hc#1{\indent\hbox to .7cm{\hfill}\hbox to 1.1cm{\n {#1}\hfill}}
\def\hd#1{\indent\hbox to 1.8cm{\hfill}\hbox to 1.4cm{\n {#1}\hfill}}
\def\n{\noindent}
\def\no{\noindent}

\def\undbib{\underbar {\hbox to 2cm{\hfill}}, }
\def\lsim{{\buildrel <\over \sim}}

\def\mod#1{ ( {\rm mod}\, #1 )}

\def\half{{1\over 2}}

\def\third{{1\over 3}}

\def\PLB{ {\it Phys.~Lett.~}{\bf B}}

\def\NPB{ {\it Nucl.~Phys.~}{\bf B}}

\def\PRD{ {\it Phys.~Rev.~}{\bf D}}
\def\PRL{ {\it Phys.~Rev.~Lett.~}{\bf }}

\def\ZPC{ {\it Zeit.~Phys.~}{\bf C}}

\def\MPLA{ {\it Mod.~Phys.~Lett.~}{\bf A}}

\def\IJMPA#1#2#3{Int.~J.~Math.~Phys.~{\bf A#1} (#2) #3}

\def\MPLA#1#2#3{ Mod.~Phys.~Lett.~{\bf A#1} (#2) #3}

\def\NPB#1#2#3{Nucl.~Phys.~{\bf B#1} (#2) #3}
\def\NPBa#1#2#3{Nucl. Phys. {\bf B#1} (#2) #3}

\def\PLB#1#2#3{Phys.~Lett.~{\bf B#1} (#2) #3}

\def\PRD#1#2#3{Phys.~Rev.~{\bf D#1} (#2) #3}
\def\PRL#1#2#3{Phys.~Rev.~Lett.~{\bf #1} (#2) #3}
\def\PRep#1#2#3{Phys.~Rep.~{\bf #1} (#2) #3}

\def\ZPC#1#2#3{Zeit.~Phys.~{\bf C#1} (#2) #3}

\fnsymbol{footnote}

\def\ham#1{#1}
\def\beq{\begin{equation} }
\def\eeq{\end{equation} }
\def\beqn{\begin{eqnarray} }
\def\eeqn{\end{eqnarray} }
\def\nolabel{\nonumber }

\def\mod#1{{\rm (mod~2)} }

\def\et{ et.~al }
\def\Tr{{\rm Tr}}
\def\VEV#1{{\langle #1 \rangle} }
\def\ord#1{{\cal{O}}(#1) }
\def\simlt{\stackrel{<}{{}_\sim} }

\def\MP{M_{\rm Pl} }
\def\MZ{M_{\rm Z} }
\def\MSU{M_{\rm susy} }

\def\Min{ $M_{\rm I}$} 
\def\MI{ M_{\rm I}} 

\def\Mrn{ $M_{\rm R}$} 
\def\MR{ M_{\rm R}} 

\def\MU{M_{\rm unif} }
\def\Mun{ $M_{\rm unif}$ }

\def\MX{ M_{\rm X} }

\def\MS{M_{\rm string} }
\def\Mst{ $M_{\rm string}$ }

\def\HSMB{HSMB }
\def\HSMBp{HSMB}
\def\HSMs{HSMs }

\def\uop{$U(1)$ }
\def\uops{$U(1)$'s }
\def\uopi{$U(1)_i$ }
\def\uopphi{$U(1)'_{\psi}$ }

\def\uopap{$U(1)'_A$}
\def\uopa{$U(1)'_A$ }

\title{\vbox{\hbox{\rightline{\rm\small UPR-0756-T, hep-th/9708023}}
\hbox{Advances in Old--Fashioned Heterotic String Model Building}}}

\author{Gerald B. Cleaver\address{Department of Physics 
and Astronomy \\ 
University of Pennsylvania, Philadelphia PA 19104-6396}}

\begin{document}
\begin{abstract}
I review findings of various research groups 
regarding perturbative heterotic string model building
in the last 12 months. 
Attention is given to recent studies of 
extra \uops and local discrete symmetries (LDS's) 
in generic string models.
Issues covered include the role of
$U(1)$'s and LDS's in limiting proton decay,  
developments in classification of models containing anomalous $U(1)$, 
and possible complications resulting from 
kinetic mixing between observable and hidden sector $U(1)$'s. 
Additionally, recent string-derived and
string-inspired models are briefly reviewed.
   
\centerline{{\it Talk Presented at SUSY '97.}} 
\end{abstract}
\maketitle

\section{Perturbative String Model Building Circa 1996-1997}

From one point of view string theory has undergone a profound paradigm 
shift in the last few years,
a shift arguably so revolutionary that the term ``string theory'' 
will most likely be replaced by some appellation (e.g. ``M''-theory
or ``F''-theory) for the current vaguely understood underlying theory. 
However, from a differing viewpoint, things have not  
significantly changed in some aspects. 
The new understanding of string dualities,
connecting the various string 
``theories'' formerly thought to be independent, 
does not appear to significantly effect physics in the weak coupling limit of
heterotic string model building (HSMB).  Thus, 
``old fashioned'' HSMB is alive and well.
Weakly-coupled heterotic strings can still play an important role in
model building.  This talk is a review of advancements 
made in \HSMB over the course of the last year.         

Section two of the talk discusses some generic  
features of perturbative heterotic string models (HSMs) whose presence 
has no field-theoretic explanation.  
In particular, the role of extra local abelian and local discrete
symmetries (LDS's) in \HSMs are reviewed. In the (quasi-)realistic \HSMs
to date, such symmetries play an important
role in constraining proton decay (and flavor changing neutral currents),
while simultaniously allowing light neutrino masses 
\cite{pati9607,faraggi9611}.
Next, progress in classification of models containing anomalous $U(1)$
are reviewed \cite{kobayashi9612}, 
including determination of anomalous $U(1)$ constraints in orbifolds.
Section two concludes with discussion of \HSMB complications 
resulting from possible
kinetic mixing between observable and hidden sector $U(1)$'s \cite{dienes9610}.
Section three briefly examines recent progress in actual {\HSMBp}. 

\section{General Aspects of Model Building}
The gauge structure of the more phenomenologically 
realistic four-dimensional \HSMs follows the general form,
\beqn
\{
SU(3)_c\times SU(2)_l\times U(1)_Y \in G&&
\nolabel\\
\times \sum_i U(1)_i \times \sum_j \Delta_j 
\}_{\rm obs}\times&&
\label{gstr}\\
\{
\tilde{G}_{\rm NA}\times\sum_k 
\tilde{U}(1)_k \times \sum_l \tilde{\Delta}_l 
\}_{\rm hid},&&
\nolabel
\eeqn 
where $\Delta_j$ denote LDS's. Minimally embedding
the SM SU(3)$_c\times$SU(2)$_l$ symmetries 
through free fermionic realization \cite{freeferm}
corresponds to realizing
$SU(3)_c\times U(1)_c\in SO(6)$ from charges of
three complex fermions and
$SU(2)_l\times U(1)_l\in SO(4)$ from those of
two complex fermions.
This embedding fixes the SM hypercharge to be 
\beqn
U(1)_Y &=& \half U(1)_l + \third U(1)_c 
\nolabel\\
       &=&
 U(1)_{I_{3R}} + \half U(1)_{B-L},
\label{hyperQ}
\eeqn  
Non-minimal embeddings involve larger numbers of fermions and can offer
degrees of freedom in the hypercharge realization. 
Differing hypercharge definitions can    
significantly alter the phenomenology of the effective field theory.

The separation between the observable and hidden sectors is generally not
absolute. 
%In particular, the wall separating the sectors must contain holes,
%if spacetime supersymmetry (SUSY) breaking originates in the hidden 
%sector and so must be passed to the observable sector. 
Indeed, most often some of the hidden sector states also carry observable 
\uopi charges.\footnote{Furthermore, three-generation 
free fermionic models exist that contain exotic 
states carrying both observable and hidden non-Abelian 
charges \cite{chaud95b}.} 
Charges carried by both hidden and observable sector
states are often said to belong to the ``shadow sector,'' Although states
with both shadow and hidden sector charge are generally present, 
quasi-realistic phenomenology requires that they have masses 
$m_{\rm shadow} \ge m_{\rm susy}$.
      
In the conventional supergravity (SUGRA) senario, supersymmetry (SUSY)
is broken in the hidden sector at a high scale
\beq 
M^{\rm hid}_{\rm susy}
%\equiv\Lambda
\sim \sqrt{\MZ\MP}\sim 10^{10~{\rm to}~12}
\,\,\,\, {\rm GeV}.
\label{kme1}
\eeq
Gravitational interactions 
(and possibly Plank-scale-suppressed non-renormalizable superpotential terms) 
then set the observable sector SUSY-breaking mass
scale,  
\beq
M^{\rm obs}_{\rm susy}\sim
(M^{\rm hid}_{\rm susy})^2/\MP\sim \MZ.
\label{msobs}
\eeq 
In the less conventional gauge-mediated
process the hidden sector SUSY-breaking scale  $\MSU^{\rm hid}$
is typically around $100$ TeV. SUSY-breaking
is carried to the observable sector via loop-factor suppression factors.

Some interesting papers focusing on various roles of the ubiquitous 
stringy \uops have appeared in the last year. In the following 
subsections I will discuss four issues in these papers.

\subsection{\uop Beyond Those Within E$_6$}

Recently Pati has presented evidence 
supporting the stringy origin of MSSM matter.  
Ref.~\cite{pati9607} demonstrates that MSSM states in any 
phenomenologically viable model must carry 
at least one local \uop or LDS  charge not
originating from within the same E$_6$ (or subgroup thereof) \cite{font8901}
that $SU(3)_c\times SU(2)_l\times U(1)_Y$ 
(henceforth denoted by $(321)$)
may be embedded into. 
Extra \uops are
required for suppression of unsafe 
%(effective) $d= 4, 5,\dots$
color-triplet mediated and/or gravity-linked proton decay operators 
generically arising in SUSY (grand)-unification. The necessity of
an additional \uop becomes even more apparent when seesaw-type 
neutrino masses are desired. While extra $U(1)$ may be invoked
ad hoc from the field-theoretic approach, the appearance of (several)
additional \uops is a natural outcome in most string models. 
 
In MSSMs, the dangerous proton decay operators arise from 
baryon and lepton number violating superpotential terms of the form
\beqn
W &=& [\eta_1 \ham{U}^c\ham{D}^c\ham{D}^c 
     + \eta_2 \ham{Q}  \ham{L}  \ham{D}^c 
     + \eta_3 \ham{L}  \ham{L}  \ham{E}^c ]
\nolabel\\
  &+& [\lambda_1 \ham{Q}  \ham{Q}  \ham{Q}  \ham{L} 
     + \lambda_2 \ham{U}^c\ham{U}^c\ham{D}^c\ham{E}^c
\label{dangw34}\\
  &+& \phantom{[}\lambda_3 \ham{L}  \ham{L}  \ham{H_2}\ham{H_2} ]/M,
\nolabel
\eeqn
where SU(2), SU(3), and generational indices are suppressed.
$M$ is a characteristic mass scale, assumed herein to be 
$\MS \approx g_s\times 5.5\times 10^{17}$ GeV, 
where the string coupling $g_s\approx 0.7$.
 
$\eta_i$ and $\lambda_j$ represent terms of generic order
and can contain 
built-in suppression factors $(\VEV{\phi}/M)^n$, where 
$\VEV{\phi}$ is a non-Abelian singlet state 
VEV.\footnote{$\VEV{\phi}$ could, in general,
be replaced by a singlet product of
non-Abelian fields, such as a condensate, 
$\langle T\overline{T}\rangle$, 
of two hidden sector vector-like fields. 
In this case, additional factors of 
$1/M$ are then implied.}  
Proton decay limits imply $\eta_1\eta_2\simlt 10^{-24}$ and
$\lambda_j/M({\rm GeV})\simlt 10^{-25}$ for $\Delta(B-L)= 0$ decays. 

While dangerous dimension four ($d=4$) 
operators from renormalizable $\eta_i$ terms in
(\ref{dangw34}) can be avoided in generic SUSY (GUT)
models either by an ad hoc discrete symmetry like $R= (-1)^{3(B-L)}$ or
by a local $U(1)_{B-L}\in SO(10)$ symmetry, neither symmetry can prevent
similar effective $d=4$ operators from non-renormalizable $\eta_i$ 
terms containing factors of $(B-L)$ violating VEVs. 
Neither are the $d=5$ operators associated with the
$(B-L)$-conserving, non-renormalizable $\lambda_1$ and $\lambda_2$ terms 
forbidden. 
These latter terms are expected in any theory associated
with gravity  unless forbidden by a new symmetry. Additionally, in SUSY GUT
models, even when sufficient suppression of $\lambda_j$ is possible,  
induced $d=5$ operators from Higgsino triplet exchange must be prevented by
doublet-triplet splitting. 
These operators can be avoided by fine-tuning in SU(5) models or
by suitable choices of large numbers of and/or high-dimensionality for 
multiplets, as is often the method in SO(10) models.
%(Additional discrete symmetries are sometimes also necessary.) 
However, such solutions do not seem compelling, 
especially when stringy constraints on allowed
representations (reps) are considered \cite{levelk}.
(See footnote 4.)

In addition,  
the standard seesaw mechanism assigning 
light masses to the (primarily) $\nu$ and heavy masses to the  
(primarily) $\nu^c$ requires breaking $(B-L)$ at 
a high intermediate scale. However, violating $(B-L)$
by two units still allows a discrete $R$-parity
to survive, thereby eliminating the $d=4$ operators. 
This occurs, for example, when the VEV of a 126 of SO(10)
gives Majorana masses to $\nu^c$. 
However, reps with dimension significantly above the adjoint's, 
such as SO(10)'s 126,
are forbidden in stringy 
level-one models \cite{levelk} and even seem unlikely to
appear at any level \cite{dienes9701}. 
Alternatively, products of VEVs for sneutrino-like fields,
$N^c$ and $\overline{N'}^c$,
 from vector-like pairs of $16_H$, $\overline{16_H}$ Higgs fields 
can play the role of the 126 and give Majorana mass to the $\nu^c$ 
from the non-renomalizable superpotential
term $16_i\cdot16_i\cdot\overline{16_H}\cdot\overline{16_H}/M$. 
Simultaneously these fields can produce another SO(10)-allowed term,
$16\cdot16\cdot16\cdot{16_H}/M$, 
which contains the dangerous 
$U^cD^cD^c\VEV{16_H}/M$ and $QLD\VEV{16_H}/M$ couplings. One should
expect $\VEV{16_H}/M\sim\ord{1/100}$,
which would predict a proton lifetime $\sim 10^{-6}$ years. 
%$\widetilde{\overline{N_R}}$ 

The appearance of an additional \uopphi from 
${\rm E}_6\rightarrow{\rm SO(10)}\times$\uopphi 
improves the situation somewhat,
but still doesn't eliminating fast proton decay.  
A 27 of E$_6$ breaks into  
$(16_{Q_{\psi}=1} + 10_{Q_{\psi}=-2} + 1_{Q_{\psi}=4})$.
The renormalizable $\eta_i$ terms in (\ref{dangw34}) are forbidden by 
$Q_{\psi}$ and $Q_T\equiv Q_{\psi} - (B-L)$ as well as by $(B-L)$.
If required for $\nu^c$ masses,
$\VEV{16_H}$ and $\VEV{\overline{16}_H}$ break
$I_{3R}$, $(B-L)$, and $Q_{\psi}$.
$Y$ and $Q_T$ remain unbroken until
the two MSSM Higgs in the $10_{Q_{T}=-2}$ or 
the singlet $1_{Q_{T}=4}$ (denoted as $\chi$) take on a VEV.
%All of the
%renormalizable $\eta_i$ terms are eliminated since they 
%carry $Q_T=4$ charge.
The lowest order non-renormalizable $\eta_i$ terms are generated by
\beqn
&&16\cdot16\cdot16 \frac{\VEV{N^c}}{M}
                   \frac{\VEV{10_H}\VEV{10_H}}{M^2},
\label{tsnrm1}\\
&&16\cdot16\cdot16\frac{\VEV{N^c}}{M}
                  \frac{\VEV{\chi}}{M},
\label{tsnrm2}
\eeqn
which if $\VEV{10_H},\VEV{\chi}\sim\ord{1\,\,{\rm TeV}}$ yields 
$\eta_1\eta_2\ll 10^{-24}$. 

Although the lowest order $\lambda_j$ terms are forbidden by 
both $Q_{\psi}$ and $Q_{T}$,
the problem of triplet exchange exists. 
An effective mass
term $\VEV{\chi}H_3 {H'}_{3^*} + {\rm hc}$ for the Higgs triplets
$H_3, {H'}_{3^*}\in 10_H$
breaks $Q_{\psi}$ and $Q_{T}$ by four units.
Exchange of these triplets would again induce
dangerous effective $d=5$ operators, with $\lambda_j$ above
acceptable limits for $\VEV{\chi}\sim\ord{1\,\,{\rm TeV}}$.   

Alternatively, Pati argues that the extraodinary stability of the proton 
should emerge as a compelling feature of a model, 
resulting from symmetries endemic to the underlying theory. 
The extra, ubiquitous \uops in heterotic string models
can often form such symmetries. 
A typical example is the model presented in 
\cite{faraggi9201}. This $(321)$ model contains a total of eight
local Abelian symmeties beyond $U(1)_{I_{3R}}$ and $U(1)_{B-L}$. 
The first three of these,
$U(1)_{i=1,2,3}$ are generational symmetries. 
Specifically, all $i^{\rm th}$ generation
MSSM states carry an extra $Q_{i}= 1/2$ charge.  
  
Both observable and hidden sector states carry 
$Q_{I_{3R}}$, $Q_{B-L}$, and $Q_{i= 1\,\,{\rm to}\,\, 6}$ charge, 
while only  hidden sector states have additional $Q_{i= 7,8}$ charge.
In this basis, $U(1)_{i= 1\,\, {\rm to}\,\, 6}$ are all anomalous: 
\beqn
\Tr U(1)_1 = \Tr U(1)_2  = \Tr U(1)_3 &=&  24
\label{tr1}\\
\Tr U(1)_4 = \Tr U(1)_5  = \Tr U(1)_6 &=& -12.
\nolabel
\eeqn
All of the anomaly can be rotated into
in a single, unique
\beqn
         U(1)'_{\rm A} &\equiv& c_A\sum_i \{\Tr\, U(1)_i\}U(1)_i
\label{rotaf1}\\
                       &=& 2\left[U(1)_1+U(1)_2+U(1)_3\right]
\label{rotaf2}\\
                       & & -\left[U(1)_4+U(1)_5+U(1)_6\right]\, ,
\nolabel
\eeqn
where $c_A$ is a normalization coefficient.
One choice of basis for the remaining non-anomalous components of the
original set of \uopi is
\beqn
U(1)'_1 &=& \phantom{2}U(1)_1 - U(1)_2,\label{nonarot1}\\ 
U(1)'_2 &=& \phantom{2}U(1)_4 - U(1)_5,\label{nonarot2}\\ 
U(1)'_3 &=& \phantom{2}U(1)_4 + U(1)_5 - 2U(1)_6,\label{nonarot3}\\ 
\hat{U}(1)_{\psi} &=& \phantom{2}U(1)_1 + U(1)_2 - 2 U(1)_3,
\label{nonarot4}\\
\hat{U}(1)_{\chi} &=& \phantom{2}\left[U(1)_1 + U(1)_2 + U(1)_3\right]
\nolabel\\
                  &+&2\left[U(1)_4 + U(1)_5 + U(1)_6\right]. 
\label{nonarot5}
\eeqn 
While none of the new stringy charges can single-handedly prevent
proton decay, Table 1 (borrowed from \cite{pati9607}) reveals  
the combination of either $(\hat{Q}_{\psi},B-L)$
or $(\hat{Q}_{\psi},\hat{Q}_{\chi})$ can accomplish this.
The charges of each pair are complimentary in their preventative roles:
when one symmetry forbids a dangerous operator, the other allows it,
and vice versa. 

Constraints from satisfactory proton stability
are minimal in this model:
Either pair of symmetries need only survive from the string scale 
down to an intermediate scale significantly below $\MS$. 
The sneutrino-like fields can take on 
VEVs up to $\sim 10^{15\,{\rm to}\, 16}$ GeV, 
and hidden sector states can condense at a scale 
as high as $10^{15.5}$ GeV.  The strongest assumption is 
that the suppression factor, $(\langle\phi\rangle/M)^n$,
from singlet state VEVs is $\leq 10^{-9}$.
 
The potential of inducing dangerous $d=5$ operators 
via exotic triplets in the infinite tower of Planck-mass states 
is entirely removed by $\hat{Q}_{\psi}$.
A true symmetry in a string theory is respected by Planck-mass states
as well. Thus, color-triplets in the massive tower can
only appear in vector-like pairs, with opposite $\hat{Q}_{\psi}$-charges.
Such mass-terms cannot induce $d=5$ proton-decay operators,       
which violate $\hat{Q}_{\psi}$. 
Note that $\hat{Q}_{\psi}$ treats generations unequally,  
as shown in (\ref{nonarot4}). 
This is in fact a characteristic necessary for $\hat{Q}_{\psi}$'s success. 
This feature distinguishes it from $Q_{\psi}$ and  
implies that $\hat{Q}_{\psi}$-like symmetries cannot arise
from within generation independent symmetries like $E_6$.
There is a significant advantage of string theory over standard
field theory regarding explanation of proton stability.
That is,
the three MSSM generations naturally arise within many classes of 
strings, while various string symmetries distinguish between these generations.

\subsection{Local Discrete Symmetries}

In addition to the standard proton decay operators appearing in the MSSM 
and the Higgs-related in a SUSY GUT, more exotic operators can appear 
in non-MSSMs unless forbidden by additional symmetries.
States that could generate such operators inevitably appear in 
realistic stringy $(321)$ models. 
A typical stringy example is exotic vector-like pairs of $SU(3)$ 
triplet/anti-triplets, $(D^{'},\bar{D}^{'})$.
Such vector pairs can produce dangerous effective $d=4$
proton-decay operators from superpotential terms of the form:
\beqn \{ \ham{U^c}\ham{D^c}\ham{\bar D}^{'},\,
         \ham{U}^c\ham{\bar D}^{'} \ham{\bar D}^{'},\,
         \ham{Q}  \ham{L}  \ham{\bar{D}}^{'}& & 
      \label{ddb}\\ 
         \ham{U}^c\ham{E}^c\ham{D}^{'},\, 
         \ham{Q}  \ham{Q}  \ham{D}^{'},\, 
         \ham{Q}  \ham{h}  \ham{D}^{'} 
\}
&\times& \left(\frac{\langle \phi\rangle}{M}\right)^n.
\nolabel
\eeqn  
The first three terms in (\ref{ddb})
are generalizations of those in (\ref{dangw34}), with
$\ham{D}^{c}\leftrightarrow\ham{\bar D}^{'}$. 

However, string models often have their own
means for disallowing such operators. 
One such method of avoidance is local discrete symmetries, 
$\Delta$ \cite{iban9101,faraggi9611,kobayashi9612}. 
Abelian $\Delta$ arise when a 
local $U(1)_{Z'}$, with charges quantized in units of $q$, is broken 
by a VEV of a state with charge $Nq$. Then a Z$_N$ LDS still 
survives and $\sum_i Q_{Z',i}$ in superpotential terms 
must equal $0$ (mod $N$).
Faraggi illustrates the role of this type of symmetry in 
\cite{faraggi9611}.\footnote{Discussion of the roles played by
stringy non-Abelian LDS's will appear in \cite{cleaver97x}.}
In the model of \cite{faraggi9301} discussed by Faraggi, 
a $Z_4$ LDS prevents, to {\it all} orders in $n$,
the appearance  of all terms in (\ref{ddb}).
The $Z_4$ LDS remains after a local $U(1)_{Z'}$
is broken by the VEV of a 
%right-handed
sneutrino carrying 
$Q_{Z'}= \half$ charge.
In \cite{faraggi9301} the $Q_{Z'}$ charges for all MSSM states
are quantized in units of $n/2$, while charges for exotic
triplets are quantized in units of $n'/4$, with  $n,n'\in Z$.    
As a result, all of the potentially dangerous exotic terms
carry $Q_{Z'}$ quantized in units of $(2n+1)/4$.
Thus, they are eliminated from the superpotential by the 
surviving $Z_4$ symmetry.
 
\subsection{Anomalous \uops}

Anomalous $U(1)$'s such as those discussed in subsection 2.1 
are typically present in compactified string models.
After rotation of all anomalies into a single $U(1)'_A$ 
via (\ref{rotaf1}),  the Dine-Sieberg-Witten (DSW) mechanism 
\cite{dsw} breaks  
$U(1)'_{\rm A}$. However, this occurs  at the expense 
of generating a Fayet-Iliopoulos (FI) D-term that must be cancelled by 
VEVs of states $\Phi_{\alpha}$ carrying the anomalous charge,
\beqn
D_{\rm A} &=& \sum_{\alpha} Q^{\alpha}_{\rm A}|\Phi_{\alpha}|^2 
+ \frac{g^2}{192\pi^2}{\rm e}^{\phi}{\rm Tr}Q_{\rm A}
\label{daf}\\
          &\rightarrow& 0\,\, .
\eeqn
Additional D- and F-flatness constraints require
\beqn
D_{i} &=& \sum_{\alpha} Q^{\alpha}_{\rm i}|\Phi_{\alpha}|^2 = 0,
\label{dif}\\
F_{\alpha} &=& \langle \partial W/\partial \Phi_{\alpha} \rangle = 0.
\label{ff}
\eeqn

The presence of an anomalous $U(1)$ has profound effects on a model.
It results in
(1) gauge rank reduction via the induced VEVs, 
(2) new non-renormalizable mass terms via the induced VEVs, 
(3) a new source to non-universality of soft-breaking scalar masses, and
(4) new contributions to the cosmological constant through the FI
D-term. Additionally, many past models have used anomalous \uops
as generation symmetries, while recently 
an anomalous \uop has been used to construct models with SUSY-breaking
\cite{duali9601,binetruy9601}. 

Until last year, determining if
a given model contained a \uopa meant computing the charge trace
for each $U(1)_i$. As a first step towards classification of 
stringy anomalous \uopap,
Kobayashi and Nakano \cite{kobayashi9612} find constraints on
the appearance of a \uopa in orbifold models of 
E$^8\times$E$^8$ class.
They show that a \uopa will be absent if there is no 
massless twisted state 
carrying both observable and hidden sector charges or
if certain types of discrete symmetries are found. 
Moreover,
a given \uopi in the observable (hidden) sector
is anomaly free if there is a hidden (observable) group $G$  such that 
all massless $G$-charged state have vanishing charge $Q_i$. 
Relatedly, no model with an unbroken E$_8$ can have an anomalous \uopap.

Another constraint is 
if a model is invariant under a Z$_2$ momentum duality
$\bar{P}_i \rightarrow -\bar{P}_i$  in the compactified direction X$_i$,
then the charges of the corresponding \uopi sum to zero (with the exception
that if $\bar{P}_i$ is a spinorial momentum, then a second, simultaneous,  
$\bar{P}_j \rightarrow -\bar{P}_j$ is needed).

As a stepping stone for a general \uopa classification procedure for 
orbifolds, Kobayashi and Nakano demonstrate  that  analysis of 
orbifolds in a Wilson-line 
%\cite{wilson} 
background can essentially
be reduced to analysis in the absence of a Wilson-line. 
Classification of \uopa in terms of modular invariant shifts in a model
is introduced. Of the five distinct classes
of shifts found, some classes are shown to lead to anomalies while others 
do not. Lastly, it is suggested that LDS's will generally remain
after breaking of the \uopa and that these LDS's apply additional
constraints on flat directions. There may also be a relationship between
the appearance of discrete $R$-symmetries and \uopa.

Classification of \uopa in free fermionic models   
is being studied by G.C. and Alon Faraggi. Translation of
orbifold findings into constraints on basis vectors and 
phase coefficients (equivalently, $k_{ij}$-matrix entries) is underway. 
 
\subsection{Kinetic Mixing Among \uopi}

Recently a new and possibly very general constraint on phenomenologically
viable string models appeared \cite{dienes9610}.
Supersymmetry breaking hidden sectors containing local 
$\tilde{U}(1)_{k}$ 
hidden gauge factors could destabilize the observable gauge hierarchy
under certain conditions.
For both SUGRA- and GUT-mediated SUSY-breaking,
kinetic mixing has the potential to raise the electroweak scale up to the
hidden-sector SUSY-scale, $M_{\rm EW}\rightarrow M^{\rm hid}_{\rm susy}$.
This can occur when a string model 
contains states that simultaneously carry both observable 
$U(1)_{i}$ (including $U(1)_Y$) 
and hidden sector $\tilde{U}(1)_{k}$ charges. 
This occurrence is usually the case, as discussions in prior subsections 
should suggest.

Such a model may contain in its pure gauge part of the Lagrangian 
a $U(1)$ ``kinetic mixing term,'' with mixing strength measured by 
the dimensionless coefficient $\chi$:
\beqn
{\cal L}_{\rm gauge} &=& 
- \frac{1}{4}   F^{\mu\nu}_{(i)} F_{(i){\mu\nu}}
- \frac{1}{4}   F^{\mu\nu}_{(k)} F_{(k){\mu\nu}}
\nolabel\\
&& 
+\frac{\chi}{2}F^{\mu\nu}_{(i)} F_{(k){\mu\nu}},
\label{lkmt1}
\eeqn 
Supersymmetrization corresponds to,
\beqn
{\cal L}_{\rm gauge}= \frac{1}{32} \int d^2\theta
(W_{i} W_{i} + W_{k} W_{k}&&
\nolabel\\ 
\phantom{{\cal L}_{\rm gauge}= \frac{1}{32} \int d^2\theta }
            - 2\chi W_{i} W_{k}),&& 
\label{lkmt2}
\eeqn 
where $W_{i}\equiv \bar{D}^2 D V_{i}$ is the chiral gauge field strength 
formed from the vector superfield $V_{i}$. Assuming that the hidden sector
$\tilde{U}(1)_{k}$ breaks when a hidden sector state with charge 
$\tilde{Q}_k$ 
takes a VEV, the pure
gauge term in the Lagrangian 
may be brought to canonical form by performing the 
shift, $V^{\mu}_i \rightarrow V^{'\mu}_i = V^{\mu}_i - \chi V^{\mu}_k$. This 
shift leads to new interactions between the observable and hidden sectors:

\no (1) observable states $\varphi$ get hidden sector charges 
proportional to their observable sector charges and couplings,
\beq
{\cal D}^{\mu}_k = \partial^{\mu} + i(g_k \tilde{Q}_k + g_i \chi Q_i) 
V^{\mu}_k,
\label{oshsc}
\eeq
\no (2) observable states couple to hidden sector gauginos $\lambda_k$,
\beq
{\cal L} = i\sqrt{2} g_i Q_i \chi \varphi^{\dagger} \tilde{\varphi} \lambda_k 
+ \phantom{\tilde{\varphi}} {\rm h.c.},
\label{oshsg}
\eeq
\no (3) the hidden sector $D_k$ term is shifted,
\beqn
D'_i &=& -g_i \sum_a Q_{i_a} |\varphi_a |^2,
\label{hsda}\\ 
D_k  &=& -g_k \sum_b \tilde{Q}_{k_b} |\Phi_b |^2 + \chi D'_i,
\label{hsda2} 
\eeqn
\no (4) the hidden sector $D_k$ term makes a contribution
to soft scalar masses $(m_a^2)$ by an amount,
\beq
(m_a^2)_{KM} = g_i \chi Q_{i_a} \langle D_k\rangle.
\label{massc} 
\eeq

Whether or not these mixing effects are significant depends on the ratio
$\eta\equiv \chi \langle D_k \rangle / M_Z^2$. $\eta \gg 1$ suggests kinetic
mixing could be the dominant messenger of SUSY-breaking and produce
a hierarchy destabilization as $M_{\rm scalar} \gg M_Z$. While 
$\eta\sim \ord{1}$ does not destablize the gauge group hierarchy, 
it still yields significant corrections to soft scalar masses.   
On the other hand, if $\eta\ll 1$ then all mixing effects are ignorable. 

Stability of a model against KM effects is assured if
hidden sector states appear as conjugate pairs for each 
$\tilde{U}(1)_{k}$
involved in hidden sector SUSY-breaking. This produces a complete 
cancellation of kinetic mixing terms at all orders, (i.e., $\chi = 0$).
Existence of such conjugate pairs 
implies a certain hidden sector Z$_2$ LDS $\tilde{\Delta}_l$.
For generic string models lacking such conjugate states, 
it is estimated that even if $\chi=0$ at \Mst at tree level,
one-loop effects should generate non-zero values at lower scales,
\beq
\frac{\chi}{g_i g_k} = \frac{b_{ik}}{16\pi^2}  
\ln \frac{M_{\rm string}^2}{\mu^2}
+ \frac{1}{16\pi^2} \Delta_{ik}.
\label{chirge}   
\eeq
The first term on the right-hand side of eq.~(\ref{chirge}) is the massless
sector contribution, where
\beq 
b_{ik} = -{\rm Str}_{\rm massless} \bar{Q}_H^2  Q_i \tilde{Q}_k,
\label{chirged}
\eeq 
with $\bar{Q}_H^2$ the spacetime helicity operator.
The second term
is an effect of the infinite tower of massive string states. Typical   
four-dimensional string models yield 
\beq
10^{-1}\lsim |\Delta_{ik}|\lsim 10^{+1}, 
\label{rangedik}
\eeq
which corresponds to
\beq
 3\times 10^{-4}\lsim \frac{\chi(M_Z)}{g_b(M_Z)} \lsim 3\times 10^{-2}.
\label{rangex1}
\eeq
Thus, seemingly one should expect $\eta(M_Z)\gg 1$ in generic string 
models following either SUGRA or gauge-mediated SUSY-breaking paths.

With respect to this estimated range of typical $\chi$-values, 
stringy three-generation 
$(321)$ models (and GUT/semi-GUT generalizations) with $N=1$ SUSY 
appear in striking contrast to the average model.
Amazingly, in generic $N=1$ $(321)$ models, 
contributions to $\chi$ cancel level-by-level, 
starting with the massless states. 
However, the authors of \cite{dienes9610} argue that, while 
cancellation is exact for tree-level masses, 
cancellation is unstable against mass splittings from GUT-breaking, 
non-zero VEVs related to the DSW mechanism, 
low-energy SUSY-breaking, and resulting RGE-flows.  
$\chi/g_k$ contributions from mass splittings resulting from these 
first four effects are estimated to be of $\ord{10^{-3}}$,
$\ord{10^{-8} {\rm ~to~} 10^{-5}}$, $\ord{10^{-8}}$, and
$\ord{10^{-15}}$, respectively. Much of these range of values 
would yield either $\eta$ in the dangerous $\gg 1$ range or in the 
phenomenologically interesting $\eta \sim 1$ range. Thus, avoiding
potentially dangerous kinetic mixing effects in quasi-realistic heterotic
string models may imply the appearance of appropriate $Z_2$ LDS's in such
models. However, further research into kinetic mixing appears necessary 
before definitive statements can be made regarding the effective low energy 
values of $\eta$ for the $(321)$ models.    

\section{Progress in Actual Model Building}

While definitions of string model building can vary, 
the practice can be grouped into two primary classes: 
string {\it derived} models and string {\it inspired} 
%or {\it based} 
models.
A string {\it derived} model 
essentially involves a top-down approach, starting with a true
string model that yields an effective $(321)$, semi-GUT, or GUT
model in the low energy limit. 
On the other hand, a string {\it inspired} $(321)$, semi-GUT, or GUT
model follow a more bottom-up approach. It is a  
field theoretic model upon which certain string-like 
constraints are imposed. Generally these constraints include
limits on the dimensions of gauge group reps\footnote{For massless 
stringy matter to appear as  
a rep $R$ of a gauge group, unitarity constraints require 
$\lambda^R \cdot \Psi\leq K$, where
$\lambda^R$ is the highest weight of $R$, 
$\Psi$ is the highest root in the gauge group, and $K$
is the level of the gauge group. 
For example only the $1$, $10$, $16$, and
$\overline{16}$ reps can appear for $SO(10)$ at level-one. 
Adjoint Higgs (and higher dimension reps) 
can only appear at $K\geq 2$ for any gauge group. 
Further, the mass of a state is a linear function of
the quadratic Casimir of the rep, while it is an inverse function
of the level-$K$.
Thus, while all SO(10) reps up through the $210$
are allowed by unitarity at level-two, only the singlet through the $54$
can be massless.}
and resolution of the apparent disparity between
the string scale $\MS \approx g_s\times 5.5\times 10^{17}$ GeV
and the MSSM unification scale $\MU \equiv 2.5 \times 10^{16}$ GeV.  
  
Of the several proposed solutions to the scale disparity,   
findings of the last few years \cite{dienes9602}
suggest stringy GUTs \cite{cleaver96a}
and/or non--MSSM states between 1 TeV and $\MU$ are 
perhaps the only feasible ones on the list 
(neglecting unknown non-perturbative effects).
The Non-MSSM approach often generates  
semi-GUT unification near \Mun  
or additional gauge factors above $\MU$. 

The following subsections present a cursory review of developments
along both the derived and inspired lines of model construction.  
Stringy GUTs of the last year generally fall into 
the derived class \cite{kaku97,cleaver97x}, 
whereas semi-GUTs \cite{deshpande9611,murayama96,rizos9702}
and (Near/Extended-)MSSM \cite{allanach9703,bachas96} tend to be of 
the more inspired nature.
 
\subsection{String Derived GUT Models}

While several groups in recent years have been pursuing 
string-derived SUSY GUTs \cite{strgut1},
a sticking point has been the difficulty 
producing three net generations
in conjunction with adjoint (or higher dimension) Higgs.
Success was finally realized last year by Kakushadze and Tye using 
$Z_6$ asymmetric orbifolds that include a $Z_3$ outer-automorphism
\cite{kaku97}.
The latter produces level-three current algebras yielding 
three-generation $SU(5)$, $SU(6)$, $SO(10)$, and $E_6$ GUTs. 
All models in this class 
contain a single adjoint (and no higher dimension reps).
Discovering these models is undoubtedly the
most significant string GUTs advancement of the past year.
Zurab Kakushadze will speak next about these models in detail.

My study of level-two and level-four\footnote{$SO(10)$ 
level-three does not have a free fermionic realization.}
free fermionic $SO(10)$ models is ongoing. 
I have found no such models with three generations yet. 
While general arguments indicate three generations is 
more difficult to produce at these levels than at level-three, 
a no-go theorem does not exist. One may eventually result 
from my search though.  
However, if level-four models with three generations are ever found,
they may be able to contain more than one adjoint Higgs,  
possibly in conjunction with 54's \cite{dienes9701}.   
  
\subsection{String Inspired Semi-GUT Models}

Four closely related SUSY semi-GUT models were presented recently
as solutions to the MSSM-vs-string unification scale disparity.
Rizos \et \cite{rizos9702} have developed  
$SU(6)\times SU(2)_l$ and $SU(6)\times SU(2)_r$ models, while 
Despande \et \cite{deshpande9611} and Murayama \cite{murayama96} 
start at the string scale one step closer to the MSSM,    
with their $SU(4)_c\times SU(2)_l\times SU(2)_r$ models.
 
The states in these models are consistent with level-one algebras.
In Rizos et.~al, the gauge couplings of
$SU(6)$ and $SU(2)_{l\,(r)}$ meet at the string scale.
$SU(6)\times SU(2)_{l\,(r)}$ 
breaks to $SU(4)\times SU(2)_{l}\times SU(2)_{r}$ 
at a scale $\MX\approx\MU$
when $(15,1)$ and $(\overline{15},1)$ reps acquires VEVs along 
$SU(4)_c\times SU(2)_l\times SU(2)_r$ singlet directions. 
At a slightly lower scale 
$\MR\sim 10^{14.6\,{\rm to}\,16.3}$ GeV, 
$SU(4)_c\times SU(2)_l\times SU(2)_r$ breaks to
$SU(3)_c\times SU(2)_l\times U(1)_Y$ 
as $(4,1,2)$ and $(\overline{4},1,2)$ reps.~acquire 
VEVs along the $SU(3)$ singlet direction.  
Below \Mrn, the only non-MSSM states
are a pair of exotic triplets with fractional electric charge.
These become massive at $10^{5\,\,{\rm or}\,\, 6}$ GeV.

In the limit of $\MX\rightarrow \MS$, one of the few differences
between the $SU(6)\times SU(2)_l$ model and the 
$SU(4)\times SU(2)_l\times SU(2)_r$ models
is the lowering of $\MR$ to 
$10^{11\,\,{\rm to}\,\, 12}$ GeV
for the latter case.
Murayama claims these models 
are the most efficient way of resolving the scale
disparity. Each exotic state in these models plays a critical
role besides simply increasing the unification scale to near $\MS$. 

\subsection{String Inspired Extended-MSSMs}

Allanach \et \cite{allanach9703} have developed a class of extended 
MSSM models \cite{martin96}
that raises the MSSM coupling unification scale 
to approximately \Mst by the addition of
extra vector-like exotic MSSM states.
Based on typical stringy anomaly cancellation requirements,
the mass scale of the exotics, $\MI$, is assumed comparable to $\MU$. 
The exotics act on the MSSM running couplings like an RGE lens
located at \Min. This lens
refocuses the gauge coupling unification point to a value 
$\MX$ near $\MS$. 
In these models,
the $(321)$ symmetry is enhanced above $\MI$ by an extra 
$U(1)_X$ generational symmetry, whose coupling merges with the
others at $M_X$.

The range of MSSM vector-like exotics considered is limited  
to those permitted by level-one algebras. 
These are  
$[(3,2)_{Y,X}, (\overline{3},2)_{-Y,-X}]$ quark-like pairs, 
$[(3,1)_{Y,X}, (\overline{3},1)_{-Y,-X}]$  anti-quark-like pairs,
$[(1,2)_{Y,X}, (1,2)_{-Y,-X}]$ lepton doublet/
Higgs-like pairs, and
$[(1,1)_{Y,X}, (1,1)_{-Y,-X}]$ singlet pairs.
Let $n_Q$, $n_q$, $n_2$, and $n_1$ represent the number of exotics
in each of these classes, respectively. 
Several different combinations of $\{ n_Q, n_q, n_H \}$
yield $\MX \approx \MS$, with the most general requirement being
$n_q > n_Q + n_2$.
The simplest combination giving $M_X \approx \MS$ 
is  $n_q= 3$, $n_Q=2$, $n_2= 0$ plus one vector-like singlet, 
$(\Theta_{X=1},\overline{\Theta}_{X= -1})$,
to break $U(1)_X$.
Acceptable Yukawa mass textures 
result in more complicated 
models containing several vector-like Higgs 
(with a compensating increase in vector-like anti-quarks). 

Assuming a higher level for $SU(3)_c$ and $SU(2)_l$  
offers an alternative answer for the scale disparity issue.  
Bachas \et and Bastero-Gil \et \cite{bachas96} show that 
$(321)$ unification can be pushed up to \Mst by the 
addition of just $SU(3)_c$ and $SU(2)_l$ adjoint scalars with
mass around $10^{14}$ GeV. Since this mass scale is comparable
to the scale at which coupling becomes non-perturbative
for gauge groups $SU(3)$ or $SU(5)$, it is suggested that
the masses of these adjoints could be related to hidden sector
gaugino condensation.
  
\subsection{$\mu$-Term and Mass Hierarchy}

General classes of string inspired solutions have also been proposed
for the $\mu$-term problem \cite{cvetic97a,keith97,kawamura97a} 
and mass heirarchy \cite{kawamura96a,allanach9610,cvetic97a,rasin97,elwood97}. 
One of the primary inspired aspects of this pursuit 
is the general appearance and critical role played by extra $U(1)$'s.
Later this morning J.R.~Espinosa will speak about related work 
here at Penn \cite{cvetic97a}. 
So I leave further discussion of this topic to him.

\section{Concluding Comments}

``Old fashioned'' heterotic string model building is definitely
in its maturation stage. It has entered an era involving  
finely detailed (re-)analysis of (past) models.
Understanding phenomenological patterns within model classes 
has received new attention, especially so regarding the
roles of extra $U(1)$'s and local discrete symmetries (LDS's) in the models.
Extra local Abelian symmetries and LDS's play vital roles in preventing 
fast proton decay and flavor changing neutral currents.
Moreover, whenever any extra \uops are anomalous,
the phenomenology of the model is drastically altered.
This has inspired recent study of conditions required for
anomalous \uops to appear. Another avenue of model analysis
has surfaced with the discovery of possible kinetic mixing between
observable and hidden sector \uops. If this mixing is 
strong enough, possibly destablization of the observable sector 
gauge hierarchy could occur. LDS's in the hidden sector offer
a means of eliminating this danger.

As a whole, string model building in the weakly coupled realm
continues to play an important role in high energy theory.
Even in the the context of the ongoing 
second string revolution, 
there is much to be gained from both the top-down and bottom-up 
stringy approaches to model building. 

\section{ACKNOWLEDGMENTS}
G.C. thanks the SUSY '97 organizing committee, especially co-chairs
Mirjam Cveti\v c and Paul Langacker,  
for the opportunity to speak at such a well organized
 and enjoyable conference.

\begin{table} 
\centering
\begin{tabular}{|c|c|c|c|c|c|c|c|}
\hline
Operators & Generation  & $Y$ & $B-L$ & $\hat{Q}_{\psi}$ &
$\hat{Q}_{\chi}$ & $\hat{Q}_{\chi}+ \hat{Q}_{\psi}$ & If  \\
 & Combinations & & & & & & Allowed \\ \hline
 & (a) All & & & & & & \\
$U^c \,D^c \,D^c , QLD^c , LLE^c$
  & except (b)
& $\surd$ & $\times$ & $\times$ & $\times$ & $\times$ &
unsafe \\ \cline{2-8}
  & (b) 3 fields from  & & & & & & \\
 & 3 different & $\surd$ & $\times$ &
  $\surd$ & $\times$ & $\times$ &
unsafe  \\
 & families & & & & & & \\ \hline
 & & & & & & & \\
$(U^c \,D^c \,D^c$ or $QLD^c)(N^c/M)$
& All & $\surd$ & $\surd$ & $\times$ & $\surd$ & $\times$
& unsafe \\
 & & & & & & & \\
$(U^c \,D^c \,D^c$ or $QLD^c) (N^c/M)\times$
& All & $\surd$ & $\surd$
& $\surd$ &  $\times$ & $\times$ & safe \\
$[(\overline{h_1}/M)^2$ or ($\phi/M)^n]$ & & & & & & & \\
 & & & & & & & \\
$(U^c \,D^c \,D^c$ or $QLD^c) (N^c/M)\times$
& Some($\dagger$) & $\surd$ & $\surd$ & $\surd$ & $\surd$ & $\surd$
& safe \\
$(T_a \overline{T}_b/M^2)^2$ & & & & & & & \\ \hline
 & & & & & & & \\
$QQQL/M$ & All & $\surd$ & $\surd$ & $\times$ & $\surd$ & $\times$
& unsafe  \\
$(QQQL/M)(\overline{N'}^{c}_\alpha/M)_{\alpha=1,2}$ & e.g.($1,2,1,3$) & $\surd$ & $\times$
& $\surd$ & $\times$ & $\times$ & unsafe  \\
$(QQQL/M)(\overline{N'}^{c}_\alpha/M)({N^c}_\beta/M)$ & All & $\surd$ & $\surd$
& $\times$ & $\surd$ & $\times$ & safe(?)  \\
 & & & & & & & \\
$(QQQL/M)(T_a\overline{T}_b/M^2)^2$ & Some($\dagger$) & $\surd$ & $\surd$
& $\surd$ & $\surd$ & $\surd$ & safe  \\
$U^c \,U^c \,D^c \,\overline{E}/M$ & All & $\surd$ & $\surd$
& $\times$ & $\times$ & $(\ast)$ & unsafe  \\
$LL\overline{h_i}\,\overline{h_i}/M$ & All & $\surd$ & $\times$
& $\times$ & & $(\ast)$ & safe  \\
 & & & & & & & \\ \hline
\end{tabular}
\caption{(Borrowed from [1].)
The roles of $Y$, $B-L$, $\hat{Q}_{\psi}$,
$\hat{Q}_{\chi}$ and $\hat{Q}_{\chi}+ \hat{Q}_{\psi}$ 
in allowing ($\surd$) or forbidding ($\times$) the
relevant $(B,L)$ violating operators. 
The mark $\dagger$ signifies that the corresponding operator is allowed if
either two of the four MSSM fields are in
generation 1 or 2 and two fields are in generation 3, with $a=1$ and $b=3$
for the hidden sector states $T_a$ and $\overline{T}_b$ ; or
all four fields are in generation 1 or 2 with $a=1$ and $b=2$. The mark
$(\ast)$ signifies that $(\hat{Q}_{\chi}+ \hat{Q}_{\psi})$ forbids
$U^c \,U^c \,D^c \,E^c /M$ for all generation combinations
except when all four fields belong to generation 3, and that it forbids
$LL\overline{h_i}\,\overline{h_i}$ in some generation combinations,
but not in others. The Higgs doublet $\overline{h_i}$ carries
generation charge $Q_i$.}
\end{table}

\end{document}